\documentclass[11pt,twoside]{article}
\usepackage{newpasp,psfig}

\usepackage{epsf}

\index{Carilli, clc}
\index{Miley, gkm}
\index{Rottgering, hjr}
\index{pentericci, lp}
\index{kurk, jk}
\index{dharris,deh}
\index{bertoldi,fb}
\index{menten,kmm}
\index{vanbreugel,wvb}

\begin{document}

\title{High redshift radio galaxies: \\
Beacons to biased hierarchical galaxy \\
formation within large scale structure}
\author{C.L. Carilli}
\affil{NRAO}
\author{G. Miley, H.J.A R\"ottgering, J. Kurk, L.Pentericci}
\affil{Leiden}
\author{D.E. Harris}
\affil{SAO}
\author{F. Bertoldi, K.M. Menten}
\affil{MPIfR}
\author{Wil van Breugel}
\affil{LLNL}


\begin{abstract}

We summarize observations of the properties of powerful radio
galaxies and their  environments from $z = 0 \rm~ to~ 5$.
These data show that some high redshift radio galaxies
inhabit regions of gas and galaxy over-densities
indicative of proto-cluster environments.

\end{abstract}




\section{Some History }

Prior to the mid-1990's, the only high redshift galaxies known were
the parent galaxies of powerful radio sources. 
Starting with Cygnus A at $z = 0.056$ in the 1950's, 
and moving to 3C 295 at $z = 0.461$ in the 1960's, then making the
jump to 4C 41.17 at $z = 3.8$ in the 1980's, powerful radio
galaxies were, by far, the highest redshift galaxies known. 
As such, radio
galaxies have provided a unique, although admittedly biased, view of
primeval galaxies and galaxy evolution (McCarthy 1993).  The last few
years has seen the discovery of populations of `normal' galaxies at
high redshift, including optically selected UV-dropout galaxies, and
submm selected dusty starburst galaxies (Steidel et al. 1998, Hughes
et al. 1998).  At first glance, these discoveries apparently diminish
the importance of high redshift radio galaxies in the general study of
primeval galaxies. More correctly, we feel these discoveries have
placed radio galaxies in their proper context among high-z
galaxy populations. High redshift radio galaxies remain the most
(optically) luminous of the known high-z galaxies, and they have
extended, high surface brightness line and continuum emission from
radio through X-ray wavelengths. Likewise, low frequency radio surveys 
provide large, relatively unbiased samples. 
Hence, powerful radio galaxies are still the most
easily and best studied of the high-z galaxy populations, and
extensive observations across the electromagnetic spectrum have
provided information that is fundamental to our understanding of
galaxy and large scale structure formation.

While most authors agree that the parent galaxies
for luminous Fanaroff-Riley Class II radio galaxies (P$_{178 \rm MHz}
\ge$ 10$^{35}$ h$^{-2}$ erg s$^{-1}$  Hz$^{-1}$)\footnote{h $\equiv \rm
{{H_o}\over{100}}$.  We use
H$_o$ = 50 km s$^{-1}$ Mpc$^{-1}$ and q$_o$ = 0.5, unless
stated otherwise.}, 
are giant elliptical galaxies, at least to $z \approx  1.5$,
there has been debate in the recent literature concerning 
the evolution of the cluster 
environments  of powerful radio galaxies from
$z = 0$ to $ z\approx 1.5$.  Hill
and Lilly (1991) concluded that there is substantial evolution in the 
environments of powerful radio galaxies over this redshift range, from 
small groups at low-z, to Abell class 0 to 1 clusters at $z >
0.5$. The recent study by Best (2000) of intermediate-z 3C sources 
seems to support this conclusion, and he uses this fact to argue that 
intermediate-z radio galaxies cannot be the precusors of the lower
redshift sources. McLure and Dunlop (2000) have recently challenged
these findings. They find low redshift luminous AGN, both radio loud
and radio quiet, typically inhabit Abell class 0 clusters, and hence
that the mean environment for powerful radio galaxies
doesn't change dramatically from $z =  0~\rm to~ 1$. They claim that:
`powerful AGN do not avoid rich clusters, but rather display a spread
in cluster environment which is consistent with being drawn at random
from the massive elliptical galaxy population.' 
They make the point that  the co-moving space density of 
clusters at low redshift, $\approx 5\times10^{-5}$ h$^3$ Mpc$^{-3}$,
is comparable to that of powerful AGN at $z = 2.5$, with the
interesting implication that all cluster dominant elliptical
galaxies were active at $z \approx 2.5$ (West 1994). 

The study of the low-to-intermediate redshift
evolution of luminous radio galaxies is ham-strung by
the paucity of such sources at 
low-z. Figure 1 shows rest-frame P$_{178 \rm MHz}$ versus redshift for a
number of samples of radio galaxies. The 3C samples shows the clear 
Malmquist bias inherent in flux limited samples. Subsequent studies 
of lower flux density sources have filled-in the
distribution at higher redshifts by revealing lower-luminosity 
sources. However, there is a clear lack of high-power sources at $z
\le 0.5$. This is simply a consequence of our universe, and cannot be
overcome with larger samples. 

\begin{figure}
\hskip 0.5in
\psfig{figure=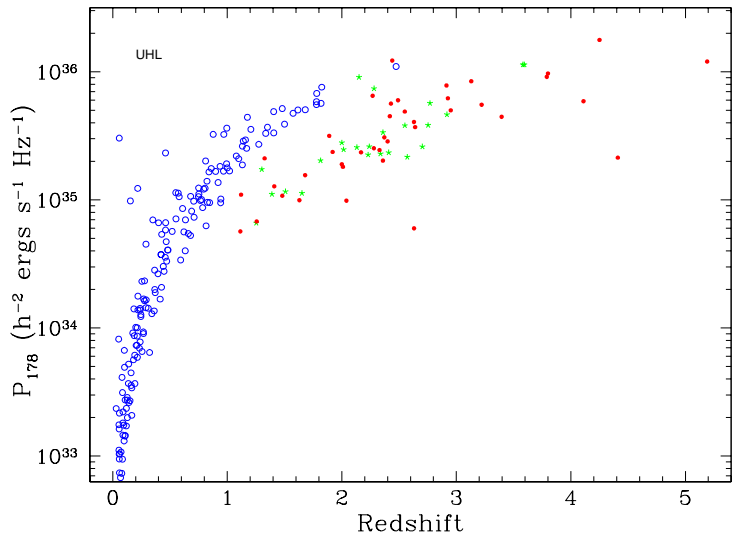,angle=-90,width=4in}
\caption{The rest-frame P$_{178 MHz}$ values versus redshift for
the 3C sample (open circles), and fainter samples used by the 
Leiden group (filled circles and stars). 
}
\end{figure}

\section{The Leiden Sample}

Through pain-staking optical identification of sources from low
frequency radio samples, a number of groups, most notably those at
Leiden, Oxford, and Berkeley, have expanded the size and redshift
range over  which radio galaxies have been detected. The current
record is  $z = 5.2$ (van Breugel et al. 1999a), and there are
150  sources known at $z > 2$. Many authors have called attention to 
the interesting parallel between the rapid increase with redshift
in the space density of  powerful radio galaxies and other
luminous AGN, with that of the cosmic star formation rate. Both show
more than an order of magnitude increase in co-moving space density
from $z = 0$ to 2 (Dunlop et al. 2000, Boyle and Terlevich 1998, Blain 
et al. 1999).  

\begin{table}
\caption{The Evolution of Powerful Radio Galaxies}
\begin{tabular}{lll}
\tableline
~ & $0 < z < 1.5$ & $ 2 < z < 5.2$  \\
\tableline
{\bf Radio} & 100 kpc, classical doubles & 10 kpc, irregular \\ 
{\bf mm to IR} & nuclear emission & distributed sources \\
~ & L$_{FIR} \le 10^{12}$ L$_\odot$ &  L$_{FIR} \ge 10^{13}$ L$_\odot$\\ 
{\bf Near IR} & giant ellipticals, old stars 
& multiple knots \\
~ & tight $K-z$ relation & weaker $K-z$ relation \\
{\bf Blue} & highly aligned with radio
& aligned with radio plus \\
~ & ~ &  off-axis knots \\
~ & $\ge 10\%$ polarized & $\le 2\%$ polarized \\
~ & ~ & stellar absorption lines \\
{\bf Ly-$\alpha$ halos} & 10 kpc, 10$^8$ M$_\odot$ & 100 kpc, 10$^9$
M$_\odot$ \\
~ & ~ & Ly-$\alpha$ self-absorption \\
{\bf Cluster} & Abell class 0 to 1 & proto-clusters? \\
~ & X-ray atmospheres & irregular X-rays \\
\tableline
\tableline
\end{tabular}
\end{table}

Table 1 gives a broad-brush summary of the properties of powerful
radio galaxies at $z \approx 1$ compared with those at $z > 2$.  While
all of the statements are defensible, some are based on observations
of just a few sources (or even one!), and remain to be verified. 

In the near-IR, the host galaxies change from well defined giant
elliptical galaxies at low-z with a fairly tight $K-z$ relation, to
spatially more complex systems at high-z,
with a larger scatter in the $K-z$
(van Breugel et al. 1999b).  
Radio galaxies at  $z \approx 1$ also show
evidence for old stars, suggesting a formation redshift $z > 3$
(Dunlop et al. 2000). 
The blue light shows a general alignment
with the radio source axis at all redshifts, but high redshift sources
also show blue knots of emission well away from the radio source
axis. Also, at low-z the blue emission is typically highly polarized,
while at high-z the emission is less polarized, or even unpolarized, 
and one source shows stellar
absorption features (Dey et al. 1997). The implication is that at
low-z the blue emission is dominated by scattered light from the
obscured nucleus, while at high-z star forming regions make a
significant contribution. Diffuse Ly-$\alpha$ emission is seen at both
low and high-z, but the Ly-$\alpha$ halos are typically larger, and
more luminous at high-z (Baum and McCarthy 2000).
Self-absorption of the Ly-$\alpha$ emission is also more prevalent in
high redshift sources, indicating large amounts of cooler  HI
(van Ojik et al. 1997, Debreuck et al. 2000).

The radio source
morphologies change from small, irregular systems at high-z to large
classical doubles at low-z. Extreme values of Faraday rotation
of the polarized radio emission are observed at both low-z and
high-z. Pentericci (1999) has argued that the fraction of sources with
large rotation measures (RMs) increases with redshift. The origin of
large RMs in low-z systems is thought to be magnetized cluster
atmospheres, with fields $\approx 10\mu$G ordered on scales of 10's of
kpc. The location of the Faraday screen in high-z sources remains an
open question. 

\section{Dust emission}

Powerful radio galaxies at low-z 
typically have IR luminosities $< 10^{12}$ L$_\odot$ year$^{-1}$, 
some of which can be ascribed to synchrotron radiation from the
AGN  (Golombek et al. 1988).
High-z radio galaxies show dramatically different behavior in the
rest-frame IR. 

Archibald et al. (1999) have shown a rapid evolution with redshift 
in the rest-frame IR luminosity of radio galaxies, with the median
value increasing as $(1 + z)^{3.5}$
to $\ge 10^{13}$  L$_\odot$ year$^{-1}$ at $z = 4$. 
The spectral energy distributions 
show that the emission is from warm dust, 
with implied dust masses $\ge 10^8$ M$_\odot$, and dust
temperatures between 50 and 100 K (Benford et al. 2000). 
Emission from CO has also been detected from a few high-z radio
galaxies, with implied gas masses $\ge 10^{10}$ M$_\odot$ 
(Papadopoulos et al. 2000).

Perhaps more importantly, imaging at submm and mm wavelengths has 
shown that the peak of the
dust and CO emission is often not centered on the AGN,
but can arise in sources distributed at distances of 1$'$ or
more from the AGN. The most dramatic example of this is 4C 41.17
at $z = 3.80$ recently imaged with SCUBA at 850$\mu$m by Ivison et
al. (2000). They detect six luminous dust emitting sources in an area
of about 4 square arcmin centered on the radio galaxy, 
including emission from the AGN itself.
Interestingly, the brightest dust emitting source is 
located at a distance of 1$'$ from the AGN. This represents an 
order-of-magnitude over-density of submm sources
relative to the field.

\begin{figure}[ht]
\hskip 0.7in
\psfig{figure=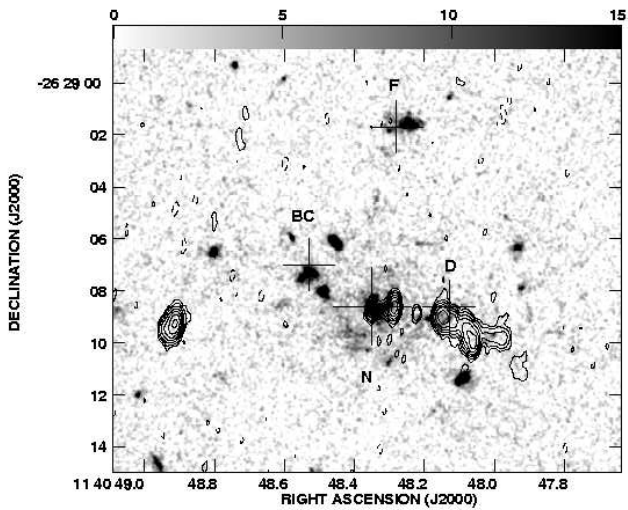,angle=-90,width=3.5in}
\caption{The greyscale shows the HST image of PKS
1138--262. The contours show the radio continuum emission at
5 GHz with 0.5$''$ resolution.   
The large cross indicates the position of the AGN, designated N.
This corresponds to the $K$-band peak intensity, and the
flattest spectrum radio component. 
The small crosses indicate off-nuclear 
Ly-$\alpha$ emitting complexes.
The IRAM 30m observations of the AGN position detected 1.5$\pm$0.5 mJy
at 250 GHz, while at the position of BC the detected flux density was 
$4.0 \pm 0.7$ mJy. At D and F the 250 GHz flux densities were $<
1.5$mJy. 
}
\end{figure}

A somewhat less dramatic, but equally telling, example of this
phenomenon is shown in Figure 2. Dust emission has been detected from
the $z = 2.156$ radio galaxy PKS 1138--262 at 240 GHz using MAMBO
(Kreysa et al. 1999) at the IRAM 30m telescope. A series of
pointed and imaging observations shows that the 
dust emission may not peak at the position of the AGN, but 
peaks about 4$''$ to the north-east of the AGN, 
at the location of a complex of Ly-$\alpha$ emitting knots
designated BC in Figure 2. 
The implied star formation rate for this complex, based on the mm
emission, is a few hundred M$_\odot$ year$^{-1}$. This is two
orders-of-magnitude greater than the 
star formation rate predicted from the rest-frame UV luminosity
(Pentericci et al. 1998), and is characteristic of the
recently discovered dust obscured, high-z starburst galaxy population
(Rowan-Robinson 1999). 

\section{A proto-cluster around PKS 1138--262}

The $z = 2.156$ radio galaxy PKS 1138--262 was chosen as the target of 
a pilot project to  search for galaxy over-densities around high
redshift radio  
galaxies for a number of reasons: (i) its extremely clumpy optical
morphology and large size that resembles simulations of forming
massive cluster galaxies (Pentericci et al. 1998), (ii) its extreme RM
and distorted radio 
morphology indicating dense local gas (Carilli et al. 1997), and (iii)
detection of extended   X-ray emission (Carilli et al. 1998).

Narrow band imaging with the VLT revealed 50 candidate Ly-$\alpha$
emitting galaxies within a $7'$ field 
centered on PKS1138--262 and in the 
redshift range of 2.137$\pm$0.037 (Kurk et al. 2000),
not including the 10 previously known
Ly-$\alpha$ emitting knots within 8$''$ of the radio galaxy 
itself (Pentericci et al 2000).
Follow-up spectroscopy of 43 candidates showed 15 of the sources
in the range z = 2.155$\pm$0.016.
There is no obvious angular sub-structure in the distribution of
the Ly-$\alpha$ emitting sources, but there is clear redshift
sub-structure, with about half the  
sources being at $z = 2.143\rm ~ to~ 2.147$,
and the other half at $z = 2.162 \rm ~ to ~ 2.166$.
The velocity dispersion of each sub-group is
about 400 km s$^{-1}$, implying a virial mass of $7 \times 10^{13}$
M$_\odot$ for each group.

The over-density of galaxies in the 1138--262 field is comparable
to that seen by Steidel et al. (1998) in their 
UV-dropout selected proto-cluster
at $z = 3.090$ (see the analysis in Pentericci et al. 2000). In terms
of large scale structure formation in the 
context of CDM models with standard cosmologies, Steidel et al. argue
that the space density of  such structures  at high redshift 
requires either a low value of $\Omega_M $ and/or a large bias parameter, 
$b$, with $b = 2$ for $\Omega_M = 0.2$, or $b = 6$ for $\Omega_M = 1$. 

\section{X-rays from PKS 1138--262}

At low-z there is a clear correlation between extreme RM radio
galaxies and X-ray emitting  cluster atmospheres (Taylor et
al. 1994). Based on this correlation, we have begun a search for
extended X-ray emission from 
high-z radio galaxies with extreme RM values. Our pilot study
has  targeted PKS 1138-262. A long exposure with ROSAT HRI detected
X-ray emission from PKS 1138--262 with a rest frame 2 to 10 keV 
luminosity of $7 \times 10^{44}$ erg s$^{-1}$, comparable to that of 
a $10^{14}$ M$_\odot$ cluster atmosphere. Unfortunately, the spatial
resolution of ROSAT was insufficient to determine the spatial
distribution of the X-ray emission, at least on scales comparable to
that expected for high surface brightness cluster core emission
$\approx 100$ kpc.

Figure 3 shows the X-ray emission from 1138--262 as seen in a 40 ksec
exposure with Chandra. The total X-ray emission (0.2 to 10 keV)
shows a point
source at the position of the AGN, plus extended emission on a scale
of about 20$''$. The extended X-ray emission
has a softer spectrum than the point source, 
contributing about 10$\%$ of the total luminosity 
in the 0.2 to 10 keV band, and 25$\%$ in
the 0.2 to 1 keV band. 

\begin{figure}
\hskip 0.5in
\psfig{figure= 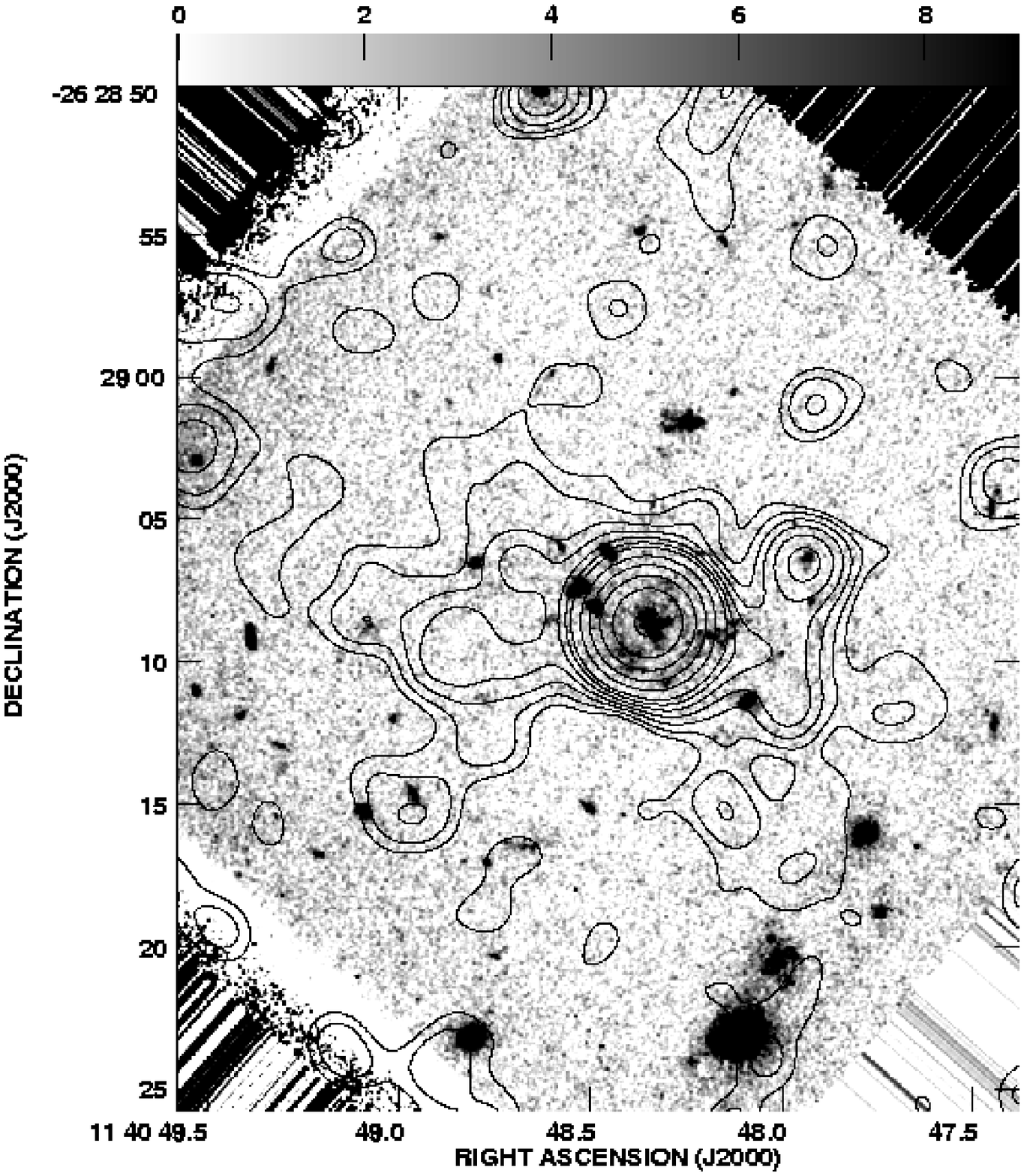,width=3in}
\psfig{figure=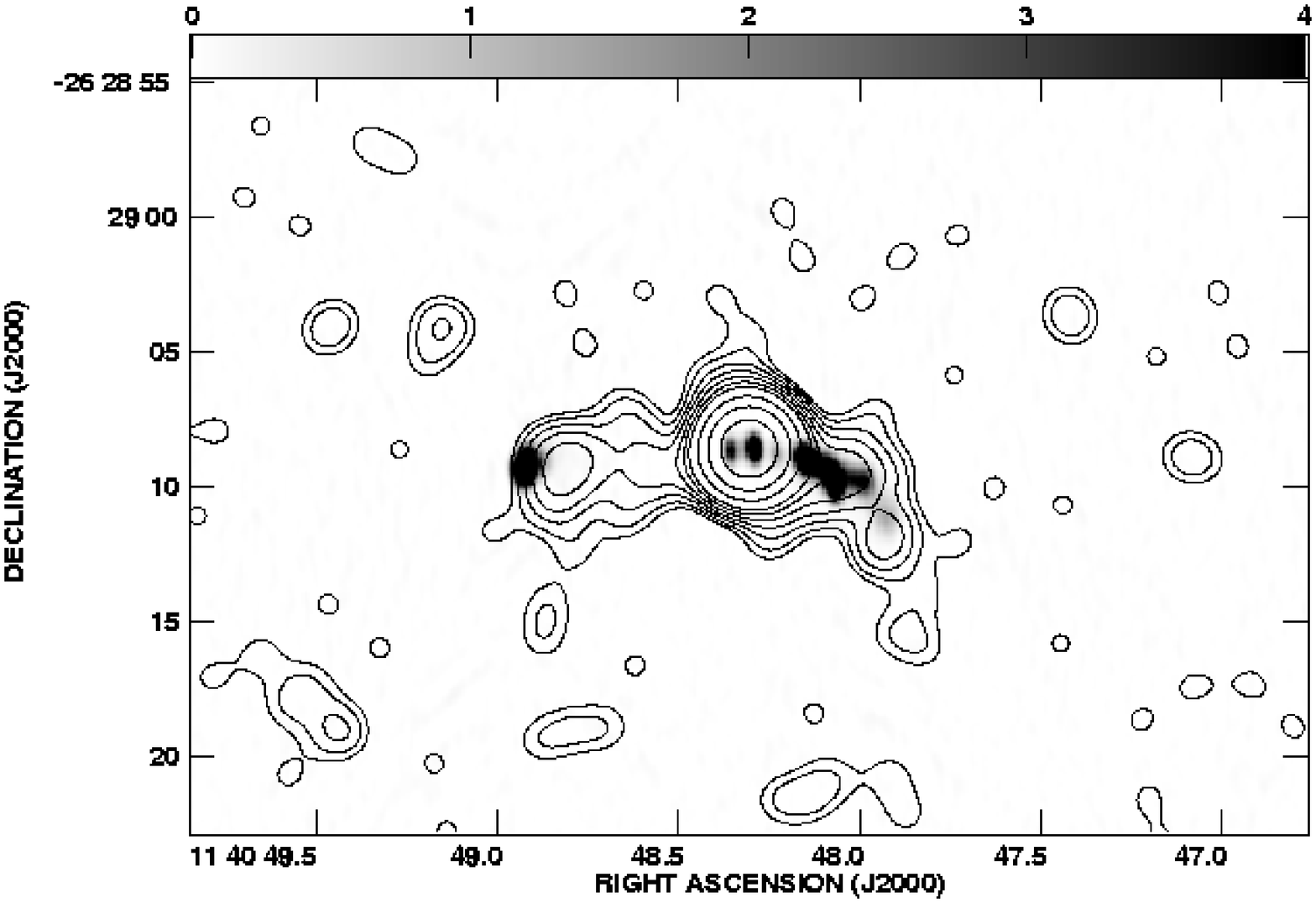,angle=-90,width=4in}
\caption{The upper frame shows a greyscale HST image of
1138--262 with the total X-ray emission observed by Chandra (0.2 to 10 
keV) as contours. The lower frame shows the soft X-ray emission
(0.2 to 1 keV) as contours, with the greyscale being the VLA 
5 GHz radio image. The Chandra images were convolved with a 2$''$
Gaussian. 
}
\end{figure}

The soft X-ray emission (0.2 to 1 keV) shows a general
correlation with the radio source in both extent and
orientation. However, the correspondence is not exact in terms of
individual high surface brightness features, most notably, the eastern
radio hot spot is situated about 2$''$ further from the nucleus than  
the local peak in the X-ray emission. Also, the X-ray emission has a
broader (transverse) distribution than the radio emission. 
Note that the relative astrometry is only accurate to about 2$''$.
However, shifting the images in order to line-up the eastern hot spot
would then move the X-ray nucleus and western extension away from the
corresponding radio structures.  Our analysis of the X-ray emission
from 1138--262  has just begun, so the following conclusions are very
tentative. 

We feel it unlikely that the X-ray emission is synchrotron radiation,
given the lack of a one-to-one correspondence with radio features,
and the fact that extrapolating the radio spectrum to the X-ray
under-predicts the X-ray flux density by more than an order of
magnitude. Inverse Compton emission is also unlikely, for similar 
spatial reasons, and the derived magnetic field is an order of
magnitude below the equipartition value. The one caveat is that, if
the radio source morphology is significantly different at low radio
frequencies relative to the high (rest-frame) frequencies observed
thus far, then inverse Compton emission might become a viable option.

One possible explanation is thermal emission from hot gas, which would 
be consistent with the large RMs observed.  The
broad-band spectrum is consistent with gas at 5 to 10 keV, although
the spectrum varies with position. 
The correlation with the radio source axis would then suggest gas 
heated by shocks driven by the expanding radio source.  
In this case, the lack of emission
from  a virialized cluster atmosphere might indicate that the 
ambient medium was at a relatively low temperature ($\le$ 1 keV) 
prior to the on-set of the radio jet.  The
observation of stronger Ly-$\alpha$ emission, and more prevalent
Ly-$\alpha$ absorption, in high redshift radio galaxies suggests a
larger supply of  cooler ambient material on scales of 10 to 100 kpc
relative to low redshift radio galaxies (van Ojik et al. 1997
Debreuck et al. 2000).  We are currently
considering the 
energetics and gas mass involved, to determine if such a model is
sensible. 


\section{High redshift radio galaxies: 
Beacons to biased hierarchical galaxy
formation within large scale structure}

Based on the observations presented above, 
we speculate that 
high redshift radio galaxies are often beacons to highly biased
hierarchical galaxy formation within large scale structure (LSS).  The
LSS will 
eventually separate from the Hubble flow and collapse to form a rich
cluster at $z \approx 1$. The individual Ly-$\alpha$ and submm-selected
galaxies will evolve into cluster spheroidal galaxies, including the
giant elliptical AGN host galaxy. 

Two question raised during the meeting are worthy of mention.
One question by Linda Sparkes 
concerns the energetics: given the 10$^{45}$ erg s$^{-1}$ 
coming from the radio AGN in these systems, could it be that all
observational characteristics are dictated by  the AGN? 
While this could be the case for regions within 50 kpc or so
of the AGN, it is clearly not the case for most of the submm and
Ly-$\alpha$ emitting galaxies distributed at distances of arcminutes
from the AGN, and, for the most part, not along the radio axis.
Also, the implied IR luminosities for the spatially distributed
submm  sources are comparable to, or larger than, the integrated
luminosity of the radio galaxy.
Hence, while the extended radio source may appear dramatic on high
quality radio images, the bolometric
luminosity of the `cluster' as a whole may
not be dominated by the radio galaxy. 
This raises the interesting possibility that the
host galaxies of high redshift radio sources may not be 
pre-ordained dominant cluster galaxies in all cases, but could become
normal cluster ellipticals, thereby alleviating the necessity for all
cluster dominant galaxies to host AGN at $z \approx 2.5$ 
(McClure and Dunlop 2000).

The second question by James Binney
concerns timescales. There is a clear correlation with
radio source size and redshift, with the median source size dropping
from 100 kpc at low-z to 10 kpc at $z = 4$ (Carilli et al. 1998). The
most recent analysis 
of this phenomenon suggests that it is due to a correlation between
radio source size, age, and luminosity: smaller sources are 
younger and more luminous (Blundell et al. 1999). As 
with the energtics discussed above, this age
discrimination  cannot dictate the characteristics of the submm
and Ly-$\alpha$ sources distributed throughout
the (proto-) cluster. Also, timescales for the radio
activity are typically 10$^6$ to $10^7$ years, while those for the
star forming galaxies are likely to be an order of magnitude
longer.  

While high-z radio galaxies are identified via  their radio
properties, ie. steep spectrum radio sources, current data
indicate that 
powerful radio galaxies and their environments represent ideal
laboratories to study the general phenomena of 
galaxy and large scale structure formation over
the widest possible range in redshifts. 
In essence, the radio source acts like a bolt of lightening
illuminating a storm -- a dramatic, short timescale event that draws
attention to a larger, even more dramatic phenomenon, 
namely the formation of a cluster of galaxies.



\acknowledgements

The National Radio Astronomy Observatory (NRAO) is
a facility of the National Science Foundation, operated under
cooperative agreement by Associated Universities, Inc..
We thank Carlos Debreuck for his tireless work on the identifications and
spectroscopic redshift measurements of many of the high redshift radio
galaxies plotted in Figure 1.  The work by W.v.B.  at the University of
California Lawrence Livermore National Laboratory was performed under
the auspices of the US Department of Energy under contract W-7405-ENG-48.

\end{document}